\documentclass[lettersize,journal,twoside]{IEEEtran}
\usepackage{amsmath,amsfonts}
\usepackage{algorithmic}
\usepackage{array}
\usepackage[caption=false,font=normalsize,labelfont=sf,textfont=sf]{subfig}
\usepackage{textcomp}
\usepackage{stfloats}
\usepackage{url}

\usepackage{verbatim}
\usepackage{booktabs}
\usepackage{upgreek}
\usepackage{bm}

\ifCLASSINFOpdf
  \usepackage[pdftex]{graphicx}
\else
  \usepackage[dvips]{graphicx}
\fi

\usepackage{graphicx}
\usepackage{xcolor}
\def\BibTeX{{\rm B\kern-.05em{\sc i\kern-.025em b}\kern-.08em
    T\kern-.1667em\lower.7ex\hbox{E}\kern-.125emX}}

\usepackage{gensymb}
\usepackage{balance}
\usepackage{tikz}
\usepackage{pgfplots}
\usepackage{verbatim}
\usepackage{tkz-tab}
\usepackage{mathptmx}
\usepackage{siunitx}
\usepackage{amssymb}
\usepackage{tabularx}
\usepackage{color}

\usepackage[hidelinks,colorlinks=true,urlcolor=black,linkcolor=blue,citecolor=blue]{hyperref}

\pgfplotsset{compat=1.18} 

\begin{document}
\bstctlcite{modify_bib}

\title{Far-Field Absolute Gain Antenna Measurements at Sub-THz Frequencies: A New Interpretation}

\author{Asad Husein,~\IEEEmembership{Graduate Student Member,~IEEE,} Kimmo Rasilainen,~\IEEEmembership{Member,~IEEE,} Juha-Pekka Mäkelä, \\Veikko Hovinen, Klaus Nevala, Aarno Pärssinen,~\IEEEmembership{Fellow,~IEEE,} and Marko E. Leinonen,~\IEEEmembership{Senior Member,~IEEE}

\thanks{Manuscript received 17 February 2026.. 
This work was supported by Business Finland RF Sampo project under Grant 2993/31/2021, by the 6G-XR project funded from the SNS JU under the EU’s Horizon research and innovation programme (Grant Number: 101096838), and by 6G Flagship (Grant Number 369116) funded by the Research Council of Finland.
\emph{(Corresponding author: Asad Husein.)}}
\thanks{The authors are with the Centre for Wireless Communications, University of Oulu, FI-90570 Oulu, Finland (e-mail: asad.husein@oulu.fi)}

}

\markboth{IEEE TRANSACTIONS ON ANTENNAS AND PROPAGATION}%
{Husein \MakeLowercase{\textit{et al.}}: Far-Field Absolute Gain Antenna Measurements at Sub-TH\MakeLowercase{z} Frequencies: A New Interpretation}

\maketitle

\begin{abstract}
The evolution of large aperture antennas and arrays at the sub-THz band (100--300\,GHz) results in traditional far-field (FF) gain measurements to require large distances due to the high frequency nature making them impractical in many laboratory environments. In the presented work, absolute antenna gain measurements are performed in localized distance clusters for commercial horn antennas in the sub-THz range of 145--170\,GHz using the three-antenna method, leveraging a theoretically derived modified FF equation along with the Friis transmission equation to enable a compact measurement setup. By applying the proposed modified FF formulation, the approach aims to redefine the FF distance by considering the combined effects of both the transmitting and receiving antennas, accounting for their aperture sizes and radiation characteristics. This allows precise gain characterization within a compact measurement footprint. The proposed theoretical model was validated through radiated measurements and simulations, demonstrating its effectiveness in this case study. Also, measurements were performed using dissimilar antenna pair combinations due to inventory constraints, a common challenge both in research and in industry. Despite the mismatches, the presented work demonstrates that reliable and sufficiently accurate measurement results can still be achieved. This highlights the practical feasibility of the compact cluster measurement technique without compromising measurement integrity. The compact setup ensures efficiency in the measurement time and cost, making it a robust solution for both research and industrial needs in sub-THz antenna characterization for applications including 6G, high frequency sensing, and imaging systems.  
\end{abstract}

\begin{IEEEkeywords}
Antenna measurements, far-field formulation, measurement uncertainty, realized gain, standard deviation, sub-THz.
\end{IEEEkeywords}

\section{Introduction}\label{intro}
\IEEEPARstart{D}{ue to} growing performance requirements, the sub-THz frequency range (e.g., at 100--300\,GHz) is receiving a lot of research focus for current fifth generation (5G) and future sixth generation (6G) wireless communication systems \cite{1}, \cite{2}. In addition to design and manufacturing, adopting sub-THz frequencies for operation creates new challenges for testing and verification. For instance, the line of sight  path loss increases with frequency and the distance between the transmitting and receiving antenna systems. Most antennas are designed to operate in the  far-field (FF), and they should therefore be tested in similar conditions as well. The well-accepted FF or Fraunhofer distance $\left(d_\mathrm{FF}\right)$ is expressed as $2D{^{2}}/\lambda$, where $D$ is the largest antenna dimension and $\lambda$ is the wavelength of operation, thus leading to impractically large FF measurement setups for sub-THz communication systems, especially with very large and/or high frequency antennas. Alternative approaches to circumvent these challenges have also been explored in recent years, such as near-field (NF) measurements, hologram and compact antenna test range methods \cite{intro1, intro2, intro3, intro4}. 

Much effort has gone into developing various measurement techniques such as \cite{gordon} that can accurately ascertain antenna gain which is one of the major performance indicators that characterizes antenna performance. The extrapolation measurement technique introduced in \cite{3} demonstrated that if a sufficient number of measurements is conducted in the distance interval of $0.2D{^{2}}/\lambda$ and $2D{^{2}}/\lambda$, gain data from the radiating NF of the antenna could then be extrapolated to the FF of the antenna. The NF-FF conversion process is accomplished using a polynomial curve fitting process, and measurements exhibit errors in gain as small as $\pm$0.11\,dB (3$\sigma$) \cite{3}. It has been further extended to compensate the effects of ground reflections encountered specially by broadbeam antennas by including phase versus distance information \cite{4}. The technique has also been utilized in \cite{wband_comp} to report an intercomparison of two W-band (75--110\,GHz) standard gain horn antennas performed across metrology facilities around the world.

Another measurement technique for determining antenna gain is the usage of the Friis transmission formula \cite{5}, \cite{6} which inherently assumes the receiving antenna to be excited by a plane wave and the separation distance between the antennas to be known. Implementations of this formula combined with the extrapolation technique include three-antenna measurements \cite{7} at an aperture-to-aperture separation distance of 5\,m and 4\,m for V-band (50--75\,GHz) and W-band (75--110\,GHz), respectively, resulting in a gain uncertainty of less than 0.1\,dB. In \cite{8}, a gain uncertainty of 0.2\,dB is achieved at the 1.0--6.0\,GHz range, which satisfies the requirement for antenna gain measurement according to the YD/T 2868-2020 communication industry standard \cite{8i}.   
 
When attempting to estimate the separation distance between the antennas, it can be difficult to determine what constitutes the reference point for the antenna. The extrapolation method generally uses the geometric physical distance between the antenna apertures. It was reported in \cite{9}, that even at an aperture-to-aperture separation as large as 16 times the $d_\mathrm{FF}$, error in measured gain was still significant, as this is related to the rather large aperture phase error intrinsic to most horn antennas. The authors of~\cite{9} further concluded that gain measurements are accurate to within $\pm$0.1\,dB at about 4 times the separation distance without any proximity correction. In \cite{morgan}, expressions for proximity correction between pyramidal and conical horn, and between dissimilar pyramidal horns are presented if the separation distances are not considerably greater than $d_\mathrm{FF}$. Another approach has been to consider the phase center of the antenna as the reference point for gain and pattern measurements. In \cite{Pivnenko}, it is shown that this would eliminate the need for proximity corrections for gain measurements at a measurement uncertainty not exceeding 0.1\,dB. However, determining the phase centers necessitates a separate measurement thereby often increasing overall measurement time and complexity \cite{Pivnenko}, \cite{odendaal}. Additional experimental results using the phase center approach are shown in \cite{12} and \cite{13}, wherein the phase centers are obtained from the simulation results.

The work done in \cite{14} assumes that radiation of a horn antenna is emitted from its amplitude center and so the separation distance is based on the displacement of the amplitude center as the receiving antenna moves towards the transmitting antenna. The gain of a pyramidal horn at 8.0--12.4\,GHz is determined using this technique with an uncertainty of $\pm$0.1\,dB. Another approach considering amplitude center as the reference point is shown in \cite{15}, where an optimization technique is applied to the Friis transmission formula in order to determine an unknown amplitude center separation distance converging to a gain value at which FF conditions are met at shorter antenna separations.    

The small wavelengths at the sub-THz band necessitate advanced measurement instrumentation such as frequency extenders, vector network analyzers, and for the most part waveguide components due to the lack of standard traceable coaxial connectors \cite{16}. Broadening the understanding of the usage of such instrumentation is critical for performing accurate gain measurements at this band. Due to the limited availability and high cost of the sub-THz components it is not uncommon within industry or research laboratories to be limited to only a pair of antennas designed for a specific frequency band. To overcome this limitation, a third antenna from a different frequency band, with some overlapping operational range, may be used in the three-antenna gain measurements. This introduces additional challenges as the differing frequency characteristics, aperture sizes, and gain profiles of the third antenna require careful calibration and compensation. 

In this article, we discuss about the challenges of using instrumentation in different measurement configurations and characterizing the limits from experimental results. Moreover, the Fraunhofer distance usually considers only the aperture size of the largest antenna, but the distance at which the FF conditions can be treated in an acceptable manner relies in practice on both antennas. When considering antennas of different bands and sizes, this may lead to an error in estimating the minimum FF distance. Hence, this work introduces a revision to the FF distance limit for aperture antennas which could be used as a starting point for such measurement configurations. Finally, path loss calculated in the Friis transmission formula must be adjusted to account for varying antenna sizes in the measurement configurations as path loss increases significantly at sub-THz frequencies. These considerations ensure that gain measurements in the sub-THz band remain reliable and adaptable to the realities of the available instrumentation. 

The outline of this paper is as follows. In Section~\ref{theory}, the theory of the three-antenna technique is presented. Here we describe the analysis of accurate path loss when using antennas of different sizes and revise the Fraunhofer equation considering the effect of both apertures. Section~\ref{measurement} is devoted to the measurement instrumentation and process. In Section~\ref{meas_analysis}, comprehensive analysis of the measurement results is presented. Finally, Section~\ref{conclusion} concludes the work.

\section{Theory of the Three-Antenna Gain Technique}\label{theory} 
The three-antenna measurement technique is a widely known measurement methodology that determines the absolute gains of antennas without the need for a reference antenna with a known gain. Assuming the antennas under test (AUTs) to be linearly polarized or having matching polarizations, three power transfer measurements are performed, one for each pair of the AUTs and thus the product of the realized gain of each antenna pair configuration can be calculated from the Friis transmission equation which serves as the foundation. The Friis transmission equation is expressed as \\
\begin{equation} 
\frac{P_r}{P_t} = G_r G_t\left(\frac{\lambda}{4\pi d}\right)^{2}
    \label{eq:1}
\end{equation}
where $P_r$ is the received power, and $P_t$ is the transmitted power at the ports of the receiving and transmitting antennas, respectively. $G_r$ and $G_t$ are the realized gains of the receiving and transmitting antennas, respectively. The free-space wavelength is denoted by $\lambda$, and $d$ is the distance between what is considered to be measurement reference point of the two antennas.

The power ratio in (1) can be represented as $|S_{21}|^2$. For each of the three-antenna pair configurations, the product of the realized gain for each pair can be determined using the measured $|S_{21}|^2$ and the free space path loss (PL) term, respectively, in the following rearranged forms \\
\begin{subequations}\label{eq:subeqns}
\begin{align}
 G_A G_B = |S_{21,{AB}}|^2\left(\frac{4\pi d_{AB}}{\lambda}\right)^{2}\label{eq:subeq1}\\
 G_A G_C = |S_{21,{AC}}|^2\left(\frac{4\pi d_{AC}}{\lambda}\right)^{2}\label{eq:subeq2}\\
 G_B G_C = |S_{21,{BC}}|^2\left(\frac{4\pi d_{BC}}{\lambda}\right)^{2}\label{eq:subeq3}
\end{align}
\end{subequations}
where ${A}$, ${B}$, and ${C}$ are the three individual antennas utilized.
\subsection{Contribution of Path Loss in Distance-Swept Measurements}
The gain measurement technique presented in this article is a modified version of the three-antenna method where, instead of measuring at a single fixed distance, a relative distance sweep is used in which one antenna moves linearly, increasing the distance between the antenna pair under measurement. By repeating measurements over multiple distances the robustness increases and helps in reducing the impact of systematic errors tied to the measurement setup.  
Aperture-to-aperture separation distance is used as the reference point of the AUTs in the measurement setup due to the convenience of a tangible and measurable distance given the small sizes of the AUTs at sub-THz frequencies. Moreover, it eliminates the complexity and does not require any additional measurement setup to identify the reference points, as discussed in Section~\ref{intro}.  

Considering an aperture-to-aperture separation, the distance between the AUTs must then be accurately accounted for in the calculation when measurements are performed over multiple distances and with AUTs of different sizes. When calculating the realized gain of a single AUT, an averaged PL term should be used to account for the differing separation distances between each of the AUT pair configurations as
\begin{equation} 
   PL = \frac{1}{3}(PL_{AB}+PL_{AC}+PL_{BC}) \approx \frac{1}{3}(3PL). 
    \label{eq:3}
\end{equation}

The modified PL term in the Friis transmission equation shown in \eqref{eq:subeq1}-\eqref{eq:subeq3} to solve gain for each AUT is then formulated as
\begin{equation} 
   PL = \frac{4\pi}{\lambda}\left(\frac{d_{AB}+d_{AC}+d_{BC}}{3}\right)[\mathrm{dB}].
    \label{eq:path_loss}
\end{equation}
\subsection{Revised Fraunhofer Equation}\label{fraunhofer-rev}
In \cite{15}, it was experimentally shown that when considering a two-antenna measurement setup, strictly speaking, the FF distance cannot be determined solely by considering the dimensions of the largest antenna. Instead, in practice, FF condition is dependent on both of the AUTs. Clearly, an equation is needed that can be used to calculate the  minimum antenna separation such that FF conditions are met considering the effects of both AUTs. 

\begin{figure}
\centering
\includegraphics[width=1\columnwidth]{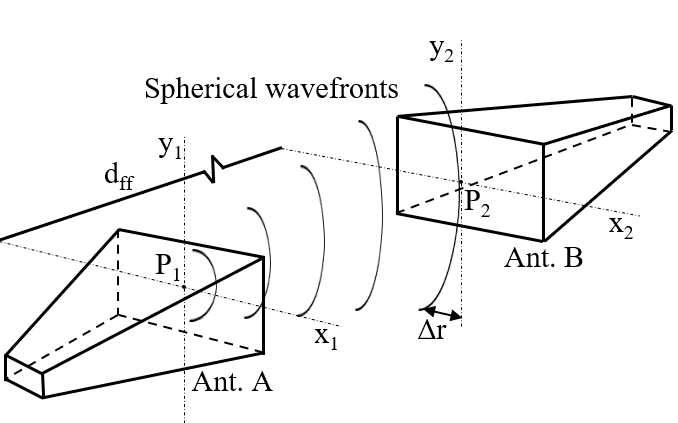}
\caption{Geometry of two similar pyramidal horn antennas whose aperture centers are aligned.}
\label{fig:geometry} 
\end{figure}

Let us consider the scenario, shown in Fig.~\ref{fig:geometry}, where two pyramidal horn antennas labelled as Ant.\,A and Ant.\,B are positioned in a direct line-of-sight configuration such that their main beams are aligned. The analysis presented could be applied similarly to the case if two dissimilar horn antennas are used. The largest dimension $D$ is the diagonal of the rectangular aperture and $d_\text{ff}$ is the axial separation between the center of the aperture $P_1$ in ($x_1$, $y_1$) plane and the center of the aperture $P_2$ in ($x_2$, $y_2$) plane.

The distance from the edge of the aperture in ($x_1$, $y_1$) plane to $P_2$ in ($x_2$, $y_2$) plane can be calculated as
\begin{equation}
\text{Distance from the edge} = \sqrt{d_\text{ff}^2 + \left(\frac{D}{2}\right)^2}.
\label{eq:4}
\end{equation}
For large $R$ (far-field assumption), we can simplify using the binomial first order approximation for square root as
\begin{equation}
\sqrt{d_\text{ff}^2 + \left(\frac{D}{2}\right)^2} \approx d_\text{ff} + \frac{\left(\frac{D}{2}\right)^2}{2d_\text{ff}}.
\label{eq:5}
\end{equation}
Spherical wavefronts that have originated from the center and the edge of the antenna aperture will have a geometric path difference when they reach point $P_2$. This path difference is what causes a phase difference between the wavefronts resulting from either constructive or destructive interference. In literature, a well-accepted FF criteria is that a maximum deviation of \(22.5^\circ\) ($\pi$/8) in the phase difference of the wavefront across the aperture of the antenna can be considered to be almost planar, indicating that FF conditions have been met~\cite[eq. (4-44)]{balanis}. The geometric path difference ($\Delta r$) can then be estimated as
\begin{subequations}\label{eq:path_difference}
\begin{align}
\Delta r &= \text{Distance from the edge} - \text{Distance from the center} \label{eq:path_difference1}\\
\Delta r &= \left(d_\text{ff} + \frac{\left(\frac{D}{2}\right)^2}{2d_\text{ff}}\right) - d_\text{ff} \approx \frac{D^2}{8d_\text{ff}}.\label{eq:path_difference2}
\end{align}
\end{subequations}
The phase error ($\phi_\text{error}$) of the spherical wavefront is related to $\Delta r$ by
\begin{equation}
\phi_\text{error} = \frac{2\pi \Delta r}{\lambda}. \label{eq:phase_error}
\end{equation}
Substituting \eqref{eq:path_difference2} in \eqref{eq:phase_error} thus gives
\begin{equation}
\phi_\text{error} \approx \frac{\pi D^2}{4\lambda d_\text{ff}}. \label{eq:phase_error1}
\end{equation}

The expression in \eqref{eq:phase_error1} informs us of the $\phi_\text{error}$ associated with the path difference from the respective aperture of Ant.\,A, assuming that Ant.\,B in the ($x_2$, $y_2$) plane is a point source. However, Ant.\,B has a certain physical size and can, in general, not be regarded as a point source. Instead, it is sampling the wavefront and integrating the phase variations it encounters across the aperture. Since the $\phi_\text{error}$ terms originating from two antennas are independent spatial displacements, in other words, center-to-edge variations of different apertures, the path difference contribution of each antenna to overall $\phi_\text{error}$ in the FF approximation can effectively be treated as orthogonal vectors. The total phase error $\left(\phi_\text{total}\right)$ would then be a combination of uncorrelated independent errors in quadrature as
\begin{equation}
\phi_\text{total} = \sqrt{\phi_\text{error,1}^2 + \phi_\text{error,2}^2} \label{eq:phase_error_total}
\end{equation}
where $\phi_\text{error,1}^2$ and $\phi_\text{error,2}^2$ are the phase error terms originating from antennas $\text{A}$ and $\text{B}$, respectively.

This root-sum-square relationship arises naturally when considering that $\phi_\text{error}$ from Ant.\,A and Ant.\,B come from independent wavefront distortions rather than coherent phase shifts. The primary contributor to phase errors is the wavefront curvature introduced by the apertures of the antennas and since the phase errors depend on the square of the aperture size \cite{balanis}, we can substitute the expressions for $\phi_\text{error,1}$ and $\phi_\text{error,2}$ in \eqref{eq:phase_error_total} as
\begin{equation}
\phi_\text{total} = \sqrt{\left(\frac{\pi D_1^2}{4\lambda d_\text{ff}}\right)^2 + \left(\frac{\pi D_2^2}{4\lambda d_\text{ff}}\right)^2} \approx \frac{\pi}{4\lambda d_\text{ff}} \sqrt{D_1^4 + D_2^4}. \label{eq:phase_error_total1}
\end{equation}
To satisfy the acceptable phase error criteria for FF distances, it is required that the condition of $\phi_\text{total}$ $\leq$ $\pi/8$ is met which on solving would give $d_\text{ff}$ as
\begin{equation}
d_\text{ff} = \frac{2}{\lambda} \sqrt{D_1^4 + D_2^4}. \label{eq:fourth_oder_distance}
\end{equation}

Although the expression in \eqref{eq:fourth_oder_distance} is a more physically accurate representation of truly independent phase distortions due to antenna apertures, following a second order dependence rather than fourth order is a more practical approximation widely accepted in antenna theory. It simplifies calculations and aligns with empirical observations that a second order expression adequately predicts FF conditions for most practical antenna configurations \cite{balanis}. 

Thus, following practical approximations, the combined aperture size contributing quadratically to the far-field phase error results in a revised separation distance ($d_\text{ff,Rev}$) corresponding to minimum FF distance as
\begin{equation}
d_\text{ff,Rev} = \frac{2\left(D_1^2 + D_2^2\right)}{\lambda}. \label{eq:revised_FF}
\end{equation}

The approximations needed to go from \eqref{eq:fourth_oder_distance} to \eqref{eq:revised_FF} are shown rigorously in Appendix \ref{ff_approximation}.
Based on \eqref{eq:path_difference2}, it is to be understood that a wavefront travelling from the edge of the aperture of one antenna traverses an extra distance of $D^2/8d_\mathrm{ff}$ to the center of the aperture of a second antenna. In the worst case offset sense, the maximum path difference across the combined link would be adding the contribution from both antenna apertures as
\begin{equation}
\Delta r_\text{max} = \frac{D_1^2 + D_2^2}{8d_\mathrm{ff}}. \label{eq:phase_error2}
\end{equation}

Hence, the total phase deviation in the worst case offset for an edge-to-edge wavefront using \eqref{eq:phase_error} is
\begin{equation}
\Delta \phi_\text{max} = \frac{\pi(D_1^2 + D_2^2)}{4\lambda d_\mathrm{ff}}. \label{eq:total_phase_error}
\end{equation}

\begin{figure}
\centering
\includegraphics[width=1\columnwidth]{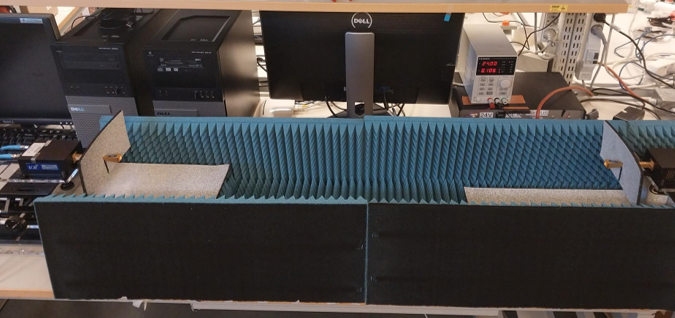}
\caption{Photo of the measurement setup and the used absorber configuration.}
\label{fig:meas_setup} 
\end{figure} 
\section{Measurement System Overview} \label{measurement}
The following subsections describe the measurement instrumentation used to build the setup, followed by the methodology of the measurement process carried out based on the understanding gained in the previous section. The compact measurement setup constructed in the RF lab at the University of Oulu, comprising of commercial pyramid absorbers fabricated for high frequency measurements is shown in Fig.~\ref{fig:meas_setup}.
\subsection{Measurement Instrumentation}
\begin{figure*}
\centering
\includegraphics[width=2\columnwidth]{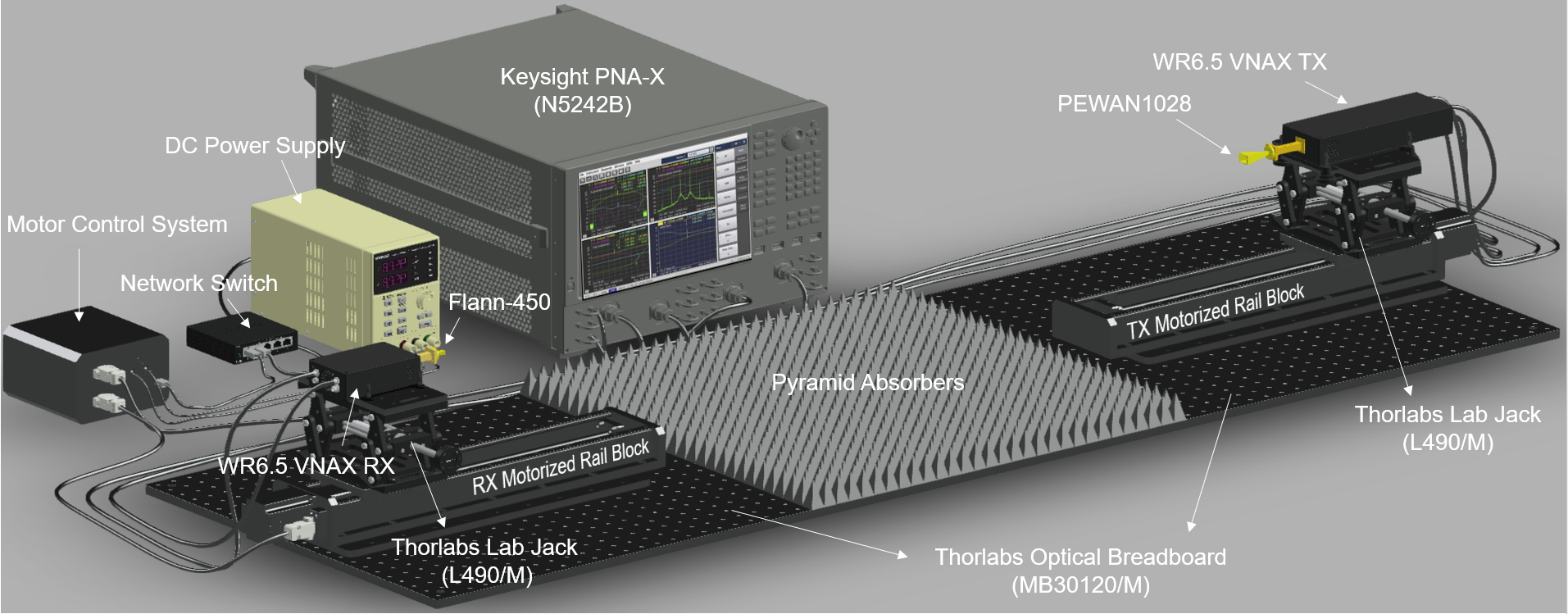}
\caption{A conceptual 3-D model of the measurement setup. Absorber side walls and flat sheet absorbers are not shown for clarity.}
\label{fig:block_diagram} 
\end{figure*}

A three-dimensional (3-D) illustration of the instrumentation utilized in the measurement setup is depicted in Fig.~\ref{fig:block_diagram}. The measurement setup consists of 
WR6.5 band (110--170\,GHz) continuous wave, transmitting (TX) and receiving (RX) frequency extenders from Virginia Diodes (VDI) Inc. They have a local oscillator (LO) multiplication factor of 12, implying that the LO input frequencies for both extenders are kept in the range of 9.17\,GHz–14.17\,GHz to cover the full range of WR6.5 band carrier frequencies. The TX and RX frequency extenders with their footplates are mounted on two separate Thorlabs L490/M lab jacks using custom 3-D printed plates. After mounting the extenders, the heights of the lab jacks were carefully adjusted to the same level with one another. 

\begin{figure}[!t]
\centering
\begin{tabular}[b]{c}
\includegraphics[width=0.95\columnwidth]{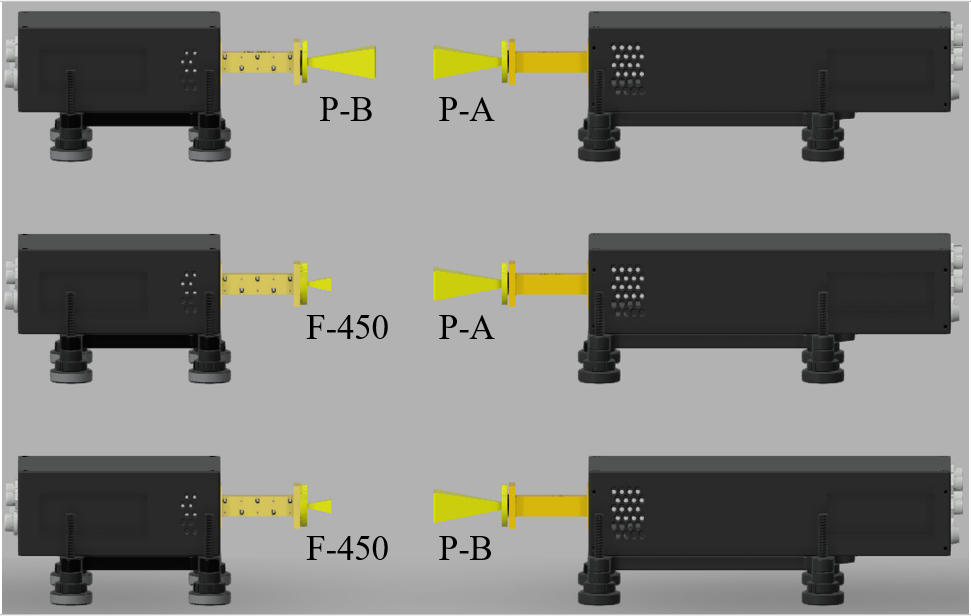} \\
\footnotesize (a) First measurement configuration
\label{fig:WR65_combos(a)}
\end{tabular} \qquad
\begin{tabular}[b]{c}
\includegraphics[width=0.95\columnwidth]{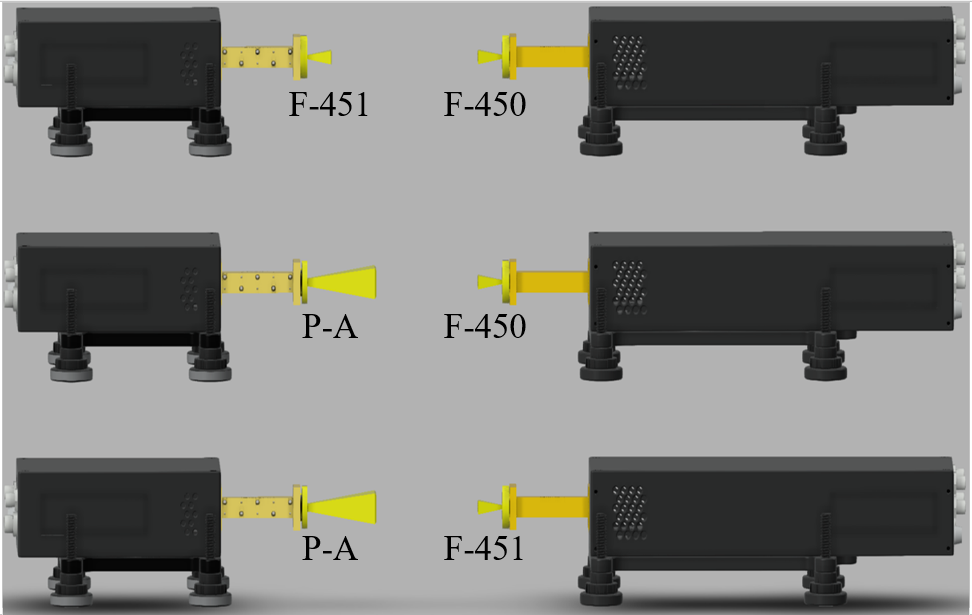} \\
\footnotesize (b) Second measurement configuration
\label{fig:WR65_combos(b)}
\end{tabular} \qquad
\begin{tabular}[b]{c}
\includegraphics[width=0.95\columnwidth]{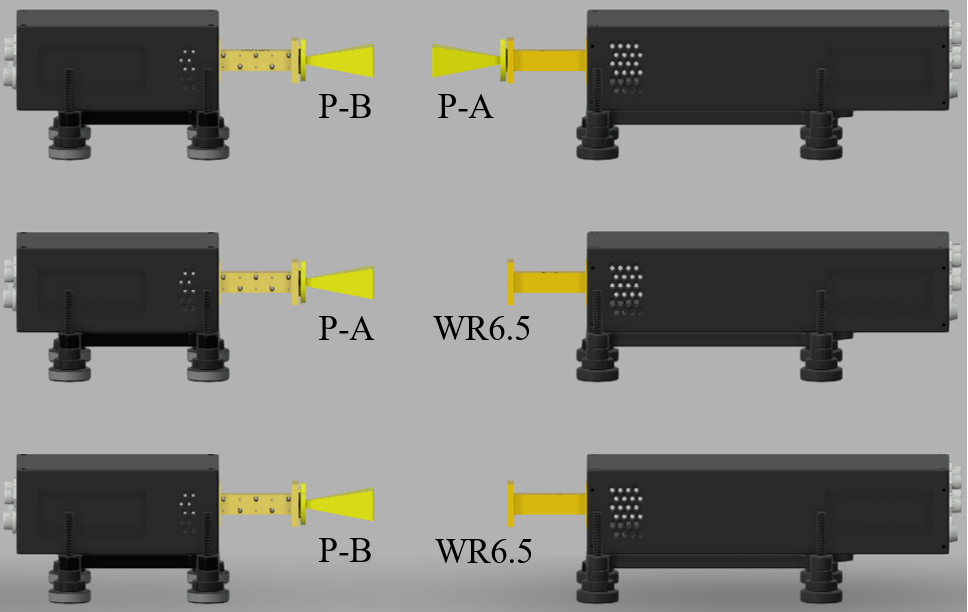} \\
\footnotesize (c) Third measurement configuration
\label{fig:WR65_combos(b)}
\end{tabular}
\caption{AUT combinations used with the three-antenna technique. The antennas are not drawn to scale. (P = PEWAN,  F = FLANN)}
\label{fig:WR65_combos}
\end{figure}
To conduct distance-sweep measurements, automation of the measurement setup is required in order to minimize measurement time among other reasons. For this, two motorized rail systems (model WMX120--350) with a single-axis translation stage is used in this setup. The translation stage is controlled over MODBUS communication via a script written in MATLAB. The travel range of the rail systems is 350\,mm having an absolute accuracy of $\pm$10\,$\upmu$m and a repeatability of $\pm$2\,$\upmu$m. The motorized rail systems are then mounted on two separate MB30120/M Thorlabs optical breadboards to ensure a solid and stable base, free from vibrations, during the measurements. The lab jacks containing the TX and RX frequency extenders are then attached to the motorized rails using a different design of custom 3-D laser cut plates. Taking into account the fact that the measurement setup comprises of several individual hardware instruments, another adjustment is performed by translating the frequency extenders to the edge point of the motorized rail and ensuring they were in the same level with respect to each other in both horizontal and vertical axes by meticulously aligning the apertures of the horn antennas to be used during the measurements. Once the required measurement range was applied to the setup, the antenna alignment was further verified by using a laser to ensure minimal deviations of the signal path from the perpendicular direction and to minimize the effects of any misalignments potentially caused by mishandling the equipment. The measurement setup is covered with WAVASORB VHP-2 \cite{absorber} pyramid absorbers. These absorbers are rated to 45$^\circ$ incidence and so the AUTs are at more than a reasonable elevation from the base to effectively suppress the shallow-angle ground reflections. In addition, the motorized rail systems are covered with flat sheet absorbers to control rail reflections, and tightly-fit flat sheet absorbers were attached around the flange portions of the AUTs to cover exposed metal surfaces of the L490/M lab jacks.

Measurements were carried out using the N5242B; a four port 26.5\,GHz performance network analyzer (PNA-X) from Keysight. Based on the available components in the lab inventory, AUTs chosen for the measurements in this study are a pair of PEWAN1028 pyramidal horn antennas from Pasternack having 25\,dBi nominal gain at the frequency range of 110--170\,GHz, and a pair of 30240 model standard gain antennas from FLANN Microwave having 20\,dBi nominal gain at the 145--220\,GHz frequency range enabling overlapping antenna measurement frequency band of 145--170\,GHz.

\subsection{Measurement Process}\label{process}
Prior to mounting the lab jacks with the frequency extenders on the motorized rail blocks, an enhanced response calibration is performed at the frequency extender waveguide output using a VDI calibration kit with open, $\lambda/4$ delay shim, and 50-$\Omega$ load standards. The calibration standards are connected to the source port, and systematic errors such as directivity, source match, frequency response reflection tracking are corrected at the source port, whereas errors like load match and frequency response transmission tracking are corrected at the receiver port.

After completing the calibration, the lab jacks with attached frequency extenders were mounted on the motorized rail blocks. Although linear translation is possible for both motorized rail blocks, it was carried out only in the TX motorized rail block as this helped to keep accurate track of the aperture-to-aperture separation distance between the AUTs. The RX frequency extender was positioned to be fixed at the edge of the RX motorized rail block in this particular setup. The measurement setup proposed in this article is currently limited to peak gain measurements, as is typically the focus of the conventional three-antenna approach. It has been shown in \cite{jaarsveld} that it is possible to perform a complete antenna characterization that includes gain and radiation pattern estimation provided measurements are performed in a spherical near field test range using transformation algorithms with higher order probe corrections.

One of the motivations of this study, as introduced in Section~\ref{intro}, is to experimentally analyze the variations in the measured results when one of the three AUTs is a specimen of overlapping frequency range and of different size and characteristics compared to the other two. The three-antenna technique is illustrated in Fig.~\ref{fig:WR65_combos}, where the first configuration uses a pair of the PEWAN1028 antennas with one sample of the FLANN antenna, the second configuration makes use of a pair of the FLANN antennas with one sample of the PEWAN1028 antenna and the third configuration uses the standalone extender waveguide with a pair of the PEWAN1028 antennas. Each individual antenna 
 was given a unique label to discern them from one another during the measurement process, hence maintaining measurement integrity. Following \eqref{eq:revised_FF}, $d_\mathrm{ff,Rev}$ distances are calculated for the AUTs at 170\,GHz, which is the highest overlapping frequency point between the two separate models. Table~\ref{table_example} presents a numerical comparison of the lower bound of the FF distance calculated at 170\,GHz between $d_\mathrm{FF}$, the formula reported by Uno and Adachi in \cite{uno} when both transmitting and receiving antennas are horn antennas $\left(d_\mathrm{ff,Uno}\right)$, the formula when considering the effect of both antenna apertures as derived in this work $\left(d_\mathrm{ff,Rev}\right)$ and the formula based on the military standard MIL-STD-449D ($d_\mathrm{ff,Mil}$) as reported in \cite{military}. The minimum distance achieved using the formula in \cite{military} is limited to a scenario where the aperture of the transmitting antenna is larger than one-tenth the aperture of the receiving antenna. To the best of the authors' knowledge, \cite{military} does not provide any further theoretical or empirical justification about this formula. 
\begin{table}[!t]
\renewcommand{\arraystretch}{1.3}
\caption{Far-field Distances Calculated At 170\,GHz Using Different Expressions}
\label{table_example}
\centering
\begin{tabular}{ccccc}
\toprule
AUT comb. & $d_\mathrm{FF}$ & $d_\mathrm{ff,Mil}$\,\cite{military} & $d_\mathrm{ff,Uno}$\,\cite{uno} & $\mathbf{d_\mathrm{ff,Rev}}$\\
\midrule
P to P & $\ge$58.3\,cm & $\ge$58.3\,cm & $\ge$233.3\,cm & $\bm{\ge}$\textbf{116.7\,cm} \\
\midrule
F to F & $\ge$8.1\,cm & $\ge$8.1\,cm & $\ge$32.3\,cm & $\bm{\ge}$\textbf{16.2\,cm} \\
\midrule
P to F & $\ge$58.3\,cm & $\ge$33.2\,cm & $\ge$109.8\,cm & $\bm{\ge}$\textbf{66.4\,cm} \\
\bottomrule
\multicolumn{4}{l}{\footnotesize{P = PEWAN,  F = FLANN}}
\end{tabular}
\end{table}

\begin{figure}
\centering
\includegraphics[]{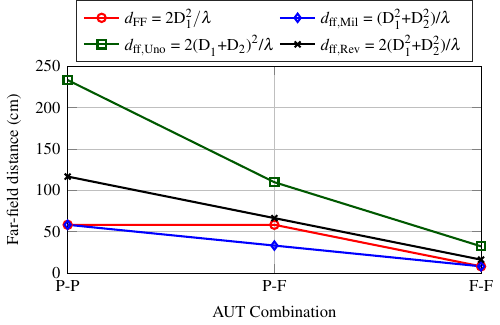}
\caption{Comparing far-field distances calculated at 170\,GHz using different expressions for various antenna model combinations. (P = PEWAN,  F = FLANN)}
\label{fig:ffs_comparison} 
\end{figure}

The data provided in Table~\ref{table_example} is represented graphically in Fig.~\ref{fig:ffs_comparison}, from which we can see that the minimum distance required as calculated using $d_\mathrm{ff,Rev}$ does not drastically increase even for the case when the pair of PEWAN1028 antennas with the larger aperture size are used. This means that measurements can still be performed using a compact setup, with some degree of accuracy, when considering the effects of both antennas using $d_\mathrm{ff,Rev}$ compared to the minimum distance obtained using $d_\mathrm{ff,Uno}$. It is also observed that the $d_\mathrm{ff,Mil}$ curve exactly overlaps with the $d_\mathrm{FF}$ curve for the cases with similar antenna models ($d_\mathrm{ff,Mil}$ reduces to $d_\mathrm{FF}$) and drops in value when dissimilar antenna models are used. 

\begin{figure}
\centering
\includegraphics[]{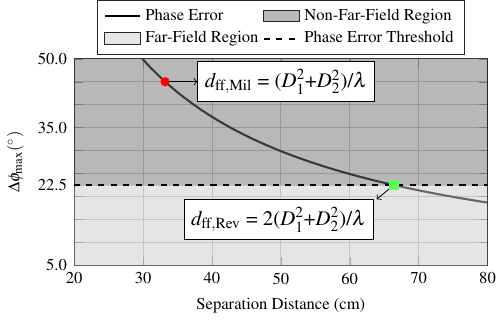}
\caption{Comparing phase error obtained at far-field distances calculated at 170\,GHz using $d_\mathrm{ff,Mil}$ and $d_\mathrm{ff,Rev}$ for the P-F combination. (P = PEWAN,  F = FLANN)}
\label{fig:phase_error} 
\end{figure}

Considering the case when dissimilar antenna models are used that satisfy the criteria in \cite{military}, e.g., the PEWAN1028-FLANN antenna models scan as illustrated in Fig.~\ref{fig:WR65_combos}(a), the minimum separation distance calculated using $d_\mathrm{ff,Mil}$ can be obtained from Table~\ref{table_example}. The phase error of the wavefront at this distance is shown in Fig.~\ref{fig:phase_error}. As expected, although $d_\mathrm{ff,Mil}$ reduces the minimum measurement distance it does not satisfy the well-accepted FF criteria as explained previously in Section \ref{fraunhofer-rev}, exceeding the phase error threshold margin by a substantial amount compared with the phase error obtained at the distance calculated using $d_\mathrm{ff,Rev}$ which satisfies the criteria.

Expanding on the results presented in Table~\ref{table_example}, compact cluster measurements (CCM) are performed at varying localized separation distances. The values of the third AUT combination in Table~\ref{table_example} can naturally be used when considering the FLANN to PEWAN1028 case as depicted in Fig.~\ref{fig:WR65_combos}(b). Cluster details are outlined in Table~\ref{table_example1}, where 151 points are measured in each cluster range at a step size of 0.2\,mm.

The measurement process is explained in detail for the the first measurement configuration, and, the same process follows for the other two configurations as well. The first measurement scan is the combination where both samples of the PEWAN1028 antennas are used. The distance between the motorized rail blocks is adjusted carefully such that the aperture-to-aperture separation distance between both the PEWAN1028 samples is 100\,cm, the starting distance in cluster\,1. Measurement data is then collected in this cluster range up to 103\,cm due to linear translation of the TX motorized rail block for 667 frequency points in the operation range.
As one of the aims of this measurement setup is compactness whilst respecting the FF boundary conditions, any errors due to mutual coupling effects can be compensated by averaging multiple measurements thereby positively affecting the overall accuracy as demonstrated numerically in \cite{Hirano2014}. For this purpose and also to confirm the repeatability of the setup, measurements are carried out five additional times, with the final dataset comprising an average of the measured results. The formulation of the measurement dataset is explained in Appendix~\ref{std_dev}. 

\begin{figure}[!t]
\centering
\begin{tabular}[b]{c}
\includegraphics{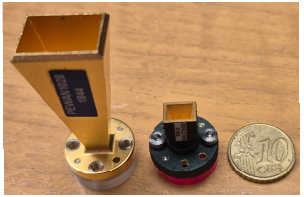} \\
\footnotesize (a) 
\end{tabular} 
\begin{tabular}[b]{c}
\includegraphics{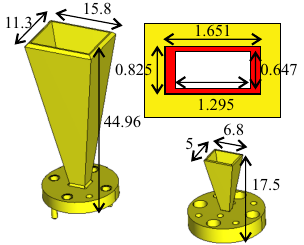} \\
\footnotesize (b) 
\end{tabular}
\caption{The used PEWAN1028 and FLANN antennas: (a) Photograph and  (b) CAD models. In (b), the red area shows the discrepancy between WR6.5 and WR5.1 waveguides. Dimensions are in millimeters.}
\label{fig:antenna_models}
\end{figure}

In order to maintain the fidelity of the measurement setup and not to introduce additional uncertainties by adjusting the position of the extender in the RX motorized rail block, an offset is introduced in the separation distances when carrying out measurements of the second and third combinations seen in Fig.~\ref{fig:WR65_combos}(a). This additional offset originates considering the smaller dimension of the FLANN antennas as seen in Fig.~\ref{fig:antenna_models}. A similar understanding can then be applied to the distances as shown in Table~\ref{table_example1}, which ensures that appropriate path loss is calculated using \eqref{eq:path_loss} for accurate gain estimation. 
\begin{table}
\renewcommand{\arraystretch}{1.3}
\caption{Measurement Distances for Different Clusters Studied}
\label{table_example1}
\centering
\begin{tabular}{cccc}
\toprule
AUT comb. & Cluster\,1 & Cluster\,2 & Cluster\,3\\
\midrule
P to P & 100--103\,cm & 120--123\,cm & 160--163\,cm \\
\midrule
F to F & 105.5--108.5\,cm & 125.5--128.5\,cm & 165.5--168.5\,cm \\
\midrule
P to F & 102.8--105.8\,cm & 122.8--125.8\,cm & 162.8--165.8\,cm \\
\bottomrule
\multicolumn{4}{l}{\footnotesize{P = PEWAN,  F = FLANN}}
\end{tabular}
\end{table}

The positional range of the motorized rail block enables us to capture data in the measurement distance range of cluster\,2 starting at 120\,cm aperture-to-aperture separation distance between the AUTs simply by modifying the positional input in the MATLAB script. In order to obtain measurement data for cluster\,3, the TX motorized rail block had to be manually shifted so as to achieve the 160\,cm aperture-to-aperture separation distance between the AUTs. 
For ensuring as accurate a boresight alignment as possible in each measurement combination for all configurations, optical laser guidance was used at both the TX and RX antennas to assist in the process. However, some misalignment due to operator intervention is to be considered within the tolerable realms of the measurement procedure due to the limitation in the translational range of the motorized rail blocks. 

The choice of measurement distances is done to satisfy the FF conditions for the largest AUT in the setup, which is the PEWAN1028 model. Cluster\,1 measurement distance implies that FF conditions are satisfied based on $d_\mathrm{FF}$ for the combination when both samples are of the PEWAN1028 antenna as given in Table~\ref{table_example1}. However, this same cluster distance lies close to the NF boundary of the PEWAN1028 antenna, if we are to consider the FF limit when calculated based on (\ref{eq:revised_FF}). Since compact measurement setups are becoming increasingly desirable at sub-THz band owing to decreased path loss and minimum measurement uncertainties, the limit given by $d_\mathrm{ff,Uno}$ is only a theoretical reference to verify if results obtained at reduced measurement distances can be classified as in the FF with some degree of accuracy. Consequently, the measurement distance at cluster\,2 is at the FF condition limit according to the revised expression of (\ref{eq:revised_FF}) for the two PEWAN1028 antenna combination. Measurements were carried out at cluster\,3 in order to have a distance which could be considered as decently in the FF while at the same time maintaining the compactness of the measurement setup.   

\section{Measurement Analysis} \label{meas_analysis}
Combining the theoretical expressions derived in Section \ref{theory} along with the measurement data obtained by carrying out the process detailed in Section \ref{measurement}, a comprehensive analysis of the measured results are presented.

\subsection{Simulation Models}
For the FLANN antennas, gain values given in the datasheet are based on calculations provided in the NRL Report 4433 with an accuracy close to approximately $\pm$0.3\,dB \cite{18}. For the PEWAN1028 antennas, a gain plot of the operational frequency range exists in the datasheet with no further information. Thus, 3-D models for both AUT specimens were constructed based on the dimensions provided, and full-wave simulations were carried out to obtain results that could be used as a comparison. Fig.~\ref{fig:antenna_models} presents the photographs and simulation models of the used antennas. It should be noted that the PEWAN1028 model is based on a CAD file availble from the vendor whereas the FLANN model has been modelled using datasheet dimensions. This accounts for the visual differences in Fig.~\ref{fig:antenna_models}(a) and (b).

The feeding waveguide of the WR6.5 frequency extender, has an internal dimension of 1.651$\times$0.825\,mm compared to the 1.295$\times$0.647\,mm internal dimension of the WR5.1 input waveguide of the FLANN antennas. This is because the FLANN antennas are designed to operate in the 145--220\,GHz band, whereas the WR6.5 frequency extenders are designed for the 110--170\,GHz band of operation. Thus, simulation results of the FLANN model by itself would not provide a good basis of comparison against the measured results. Instead, a simulation model of the FLANN antenna consisting of the WR6.5 waveguide section feeding the input port is designed. The model showing the mismatch in waveguide dimensions at the feeding port is shown in Fig.~\ref{fig:antenna_models}(b). This allows the simulation arrangement to be regarded as an emulation of the practical measurement setup, and by ignoring manufacturing flaws, the plausibility of the results obtained may be checked.

\subsection{First Measurement Configuration}
\begin{figure}[!t]
\centering
\begin{tabular}[b]{c}
\includegraphics[]{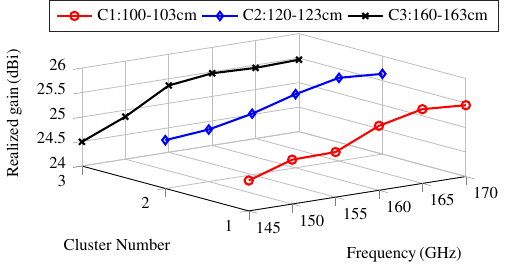} \\
\footnotesize (a) PEWAN A
\end{tabular} \qquad
\begin{tabular}[b]{c}
\includegraphics[]{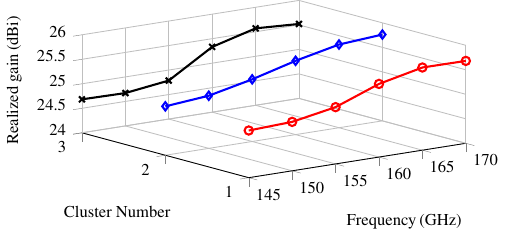} \\
\footnotesize (b) PEWAN B
\end{tabular}
\begin{tabular}[b]{c}
\includegraphics[]{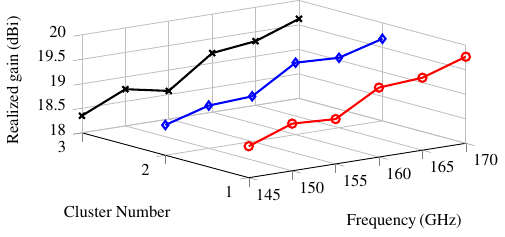} \\
\footnotesize (c) FLANN 450
\end{tabular}
\caption{Measured realized gain of the AUTs across different clusters in the first measurement configuration.}
\label{fig:3D_WR652P1F}
\end{figure}

The combination of the AUTs utilized in the measurement setup is as shown in Fig.~\ref{fig:WR65_combos}(a) for this method. Two samples of the PEWAN1028 antenna, which for the sake of differentiation have been labelled as PEWAN-A and PEWAN-B and one sample of the FLANN antenna labelled as FLANN-450 based on the vendor serial number have been used in this measurement setup. In Fig.~\ref{fig:3D_WR652P1F}, 3-D plots of the AUTs realized gain in different clusters across the overlapping frequency range is shown.
For both PEWAN1028 samples, a similar trend is observed where in cluster\,1 the realized gain is increasing with frequency having a slight drop at 170\,GHz. In cluster\,2, the increase in gain with frequency is at a more uniform rate and at cluster\,3, we notice a slight jump in the gain around the 155\,GHz point followed by small increments at the higher frequencies. Further analyzing the plots of both PEWAN1028 samples, shows the realized gain curve at cluster\,2 are pretty similar, with PEWAN-B having slightly higher measured gain values in comparison which is coming from manufacturing differences in both models. 

As explained in Section \ref{process} with regards to the choice of the measurement distances, we further extend the analysis using (\ref{eq:total_phase_error}) to verify the fulfilment of FF conditions considering the aperture effects from both AUTs. The FF distance is affected by the largest effective aperture size which would be the case when both PEWAN1028 samples are being used in the measurement for this setup. Taking the averaged value of measurement distance in each cluster, solving for (\ref{eq:total_phase_error})  then results in $\Delta \phi_\text{max}$ values of \(25.5^\circ\), \(21.6^\circ\), and \(18.5^\circ\) for clusters 1, 2 and 3, respectively. First, a clear inference from this is that FF conditions are not satisfied based on $\phi_\text{error}$ deviations in cluster\,1 when also considering the effect of the receiving antenna. Since the FF distance given by $d_\mathrm{FF}$ treats the receiving antenna as a point source, assuming measurements in the distance range of cluster\,1 to be somewhat in the FF on this basis would initially be thought as a safe assumption. However, based on the $\Delta \phi_\text{max}$ values just mentioned, this would be an incorrect assumption which thereby affirms the usage of $d_\mathrm{ff,Rev}$ for calculating FF distance when considering a two-antenna system. Secondly, as FF conditions are satisfied in the remaining two clusters, measurements in cluster\,2 could be considered as a sweet-spot, i.e besides satisfying FF criteria it also can help to minimize the measurement distance thereby leading to more compact measurement system, lesser path loss at sub-THz frequencies, and reduced effects of measurement uncertainties originating from the setup. 

In the case of the FLANN-450 antenna model, a saw-tooth like pattern is evident in all of the clusters which is consistent based on simulation results as well. Given the explicit understanding that there exists a mismatch in waveguide dimensions between the WR6.5 extender and the WR5.1 interface of the FLANN-450 antenna, small impact in results due to impedance mismatch can be expected for this particular model in either of the measurement configurations. This, however, is the purpose of these measurements in order to understand the accuracy with which gain may still be obtained within tolerable bounds given the limitations of measurement equipment.

\begin{figure}
\centering
\includegraphics[]{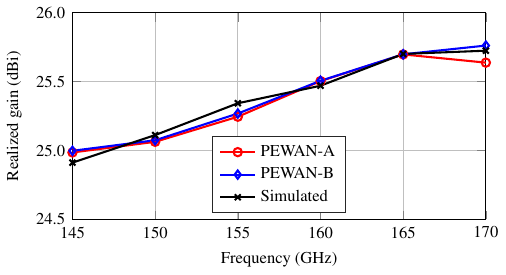}
\caption{Comparing simulated and measured realized gain within cluster\,2 for the PEWAN1028 samples in the first measurement configuration.}
\label{fig:sim_vs_PAPB} 
\end{figure}

As shown in Fig.~\ref{fig:sim_vs_PAPB}, the simulated realized gain of just the PEWAN1028 model is compared against the measured realized gain in cluster\,2 for both samples. The measured gain of the two samples differs by 0.02\,dB on average, reaching a maximum of about 0.10\,dB at the higher end of the frequency band. These obtained values are quite decent allowing for manufacturing tolerances and uncertainties arising from the measurement setup. The absolute average offset present between the simulated and measured realized gain is close to 0.06\,dB for the PEWAN-A sample and 0.05\,dB for the PEWAN-B sample which can be considered as reasonable results. 
\begin{figure}
\centering
\includegraphics[]{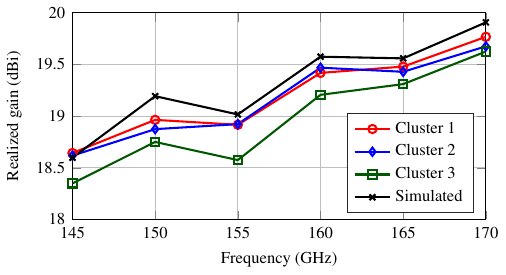}
\caption{Comparing simulated and measured realized gain across different clusters for the FLANN-450 sample in the first measurement configuration.}
\label{fig:sim_vs_FC} 
\end{figure}

The measured realized gain shown in Fig.~\ref{fig:sim_vs_FC} belongs to the FLANN-450 model compared against the simulated realized gain. The simulation model consisted of the antenna along with the WR6.5 feeding waveguide section. This way the mismatch due to waveguide dimensions would be included within the simulated realized gain results hence providing the closest reference to the actual measurement results. Taking into consideration the FF condition for a combination of the PEWAN1028 and FLANN antennas from Table~\ref{table_example}, measurements carried out at cluster\,1 itself would be quite well in the FF. The absolute average offset between simulated and measured results across the overlapping frequency range is around 0.12\,dB, 0.15\,dB and 0.34\,dB for clusters\,1, 2 and 3, respectively. The increased measurement distance at cluster\,3 results in a value having higher deviation due to increased path loss. Despite the discrepancies, the measured and simulated realized gains show a similar trend and results from clusters\,1 and 2  and they can still be considered acceptable especially from an industrial perspective.

\begin{figure}
\centering
\includegraphics[]{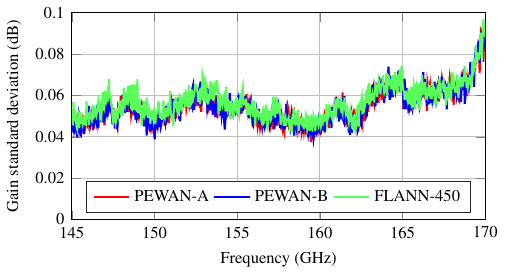}
\caption{Standard deviation in the realized gain values of all AUTs in the first measurement configuration. The standard deviation has been estimated across a population size of 150 samples.}
\label{fig:std_dev} 
\end{figure}

The standard deviation of the measured realized gain for this measurement configuration is determined using the formulation presented in Appendix \ref{std_dev}, and the results are shown in Fig.~\ref{fig:std_dev}. As can be seen in the figure, standard deviation is less than 0.06\,dB on average over 150 samples for all of the antenna models employed in the measurement setup with the maximum of about 0.09\,dB at 170\,GHz.   

\subsection{Second Measurement Configuration} \label{sec2}
As shown in Fig.~\ref{fig:WR65_combos}(b), this configuration utilizes a pair of the FLANN antennas, labelled FLANN-450 and FLANN-451 along with a single PEWAN-A antenna. From Table~\ref{table_example}, $d_\mathrm{ff,Rev}$ indicates the minimum separation distance to be around 66.4\,cm for this configuration. The purpose of this measurement configuration is to analyze the limits of accuracy for the measured gain data given the limitations of the measurement instrumentation. If we are to consider a FF distance around 70--73\,cm, from \eqref{eq:total_phase_error} taking an averaged value for distance, we would obtain wavefronts at an angle of deviation of \(20.8^\circ\) which technically satisfies the FF condition. However, we note that the PEWAN-A model is at the receiving end in both measurement runs implying that it would be able to collect more integrated energy on its wider aperture face. Thus, the wavefronts, although considered as FF, would still have a deviation enough to affect the accuracy of the realized gain of the AUT specially when comparing with simulated models. This implies the need of a tradeoff in the separation distance such that results with acceptable accuracy can be obtained without sacrificing compactness of the measurement setup. 
\begin{figure}[!t]
\centering
\begin{tabular}[b]{c}
\includegraphics[]{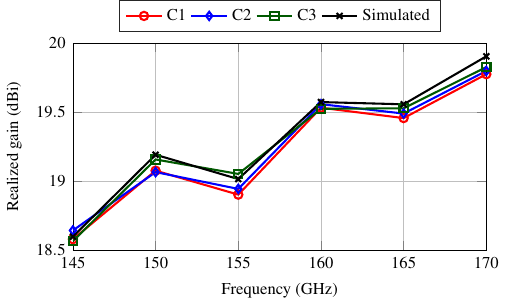} \\
\footnotesize (a) FLANN 450
\end{tabular} \qquad
\begin{tabular}[b]{c}
\includegraphics[]{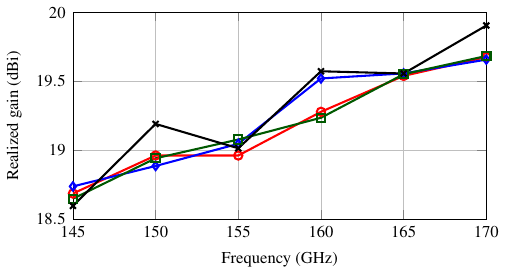} \\
\footnotesize (b) FLANN 451
\end{tabular}
\begin{tabular}[b]{c}
\includegraphics[]{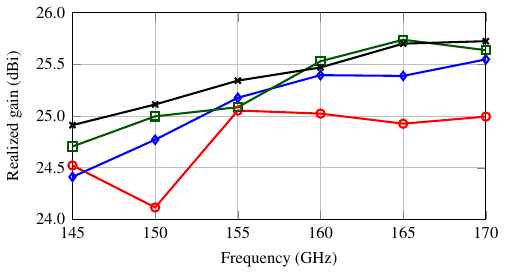} \\
\footnotesize (c) PEWAN A
\end{tabular}
\caption{Measured realized gain of the AUTs across different clusters in the second measurement configuration.}
\label{fig:sim_vs_meas_WR652P1F}
\end{figure}

On the basis of the preceding reasoning, measurements were carried out in the same cluster distances as the first measurement configuration. This allowed us reconfigurability of the existing setup without the need to rebuild for adjusted shorter separation distances, thereby reducing overall measurement campaign time. The comparison between simulated and measured realized gain across all clusters is presented in Fig.~\ref{fig:sim_vs_meas_WR652P1F}. We see that the measured realized gain in each cluster follows the trend of the simulations for the FLANN-450 sample. The absolute average offset between simulated and realized gain across the operational frequency range are found to be 0.08\,dB, 0.07\,dB and 0.04\,dB for clusters\,1, 2 and 3, respectively. For the FLANN-451 sample, absolute average difference values are found to be 0.15\,dB, 0.13\,dB and 0.16\,dB for clusters\,1, 2 and 3, respectively. From these results we can infer that as we move deeper into the FF, gain values obtained will improve due to the increasing planarity in the geometry of the wavefronts. The discrepancies in the values between the samples indicates the effects of manufacturing tolerances and errors in the FLANN-451 sample. In the case of the PEWAN-A sample, absolute average offset values of 0.60\,dB, 0.26\,dB and 0.13\,dB for clusters\,1, 2 and 3, respectively are obtained. The results obtained can safely be considered within the limits of absolute gain uncertainty value of 0.60\,dB and 0.30\,dB standard uncertainty value as specified in the technical report (TR 138 903 V15.2.0) of the European Telecommunications Standards Institue (ETSI) \cite{3gpp}. Although these numbers are limited to millimeter-wave frequencies, based on the authors knowledge there are no standards currently available for sub-THz frequencies, therefore bigger uncertainties are quite possible.
\begin{figure}
\centering
\includegraphics[]{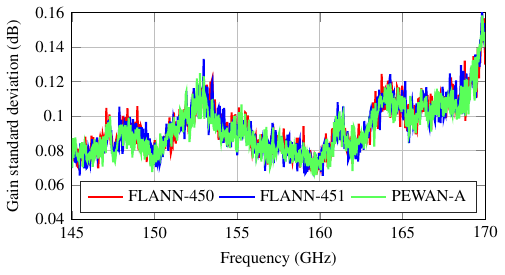}
\caption{Standard deviation in the realized gain values of all AUTs in the second measurement configuration. The standard deviation has been estimated across a population size of 150 samples.}
\label{fig:std_dev1} 
\end{figure}
Similar to Fig.~\ref{fig:std_dev}, standard deviation of the measured realized gain, calculated based on formulation presented in Appendix \ref{std_dev} for this measurement configuration, is shown in Fig.~\ref{fig:std_dev1}. Across the overlapping operating range of frequencies, a decent value of around 0.09\,dB on average is calculated for each of the antenna models used in this measurement configuration.

\subsection{Third Measurement Configuration}
In this measurement configuration we perform gain measurements using the open-ended waveguide flange of the WR6.5 frequency extender as the third AUT as shown in Fig.~\ref{fig:WR65_combos}(c). The purpose of this measurement configuration is to validate the use of frequency extenders as an AUT to obtain gain results, within acceptable bounds of precision. Although, measurements for two AUTs of the same operational frequency can be done using the two-antenna method, however, it requires the gain of at least one of the AUTs to be known beforehand which may introduce uncertainty if the reference AUT gain is not precisely characterized. In addition, the three-antenna technique inherently reduces systematic errors by cross-calibrating all three antennas, leading to a more robust gain determination. The aperture dimension of the stand alone extender waveguide is quite negligible compared to the aperture dimension of the PEWAN1028 antenna. However, the first scan as seen in Fig.~\ref{fig:WR65_combos}(c) would be the combination of both PEWAN1028 samples and so measurements would still have to be performed at the cluster distances as previously discussed.  

\begin{figure}[!t]
\centering
\includegraphics[]{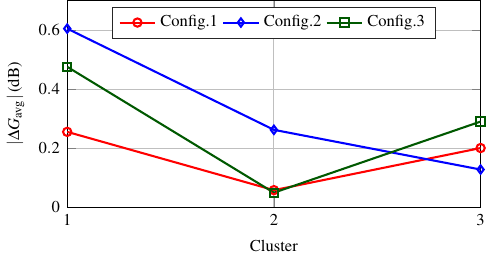}
\caption{Absolute average difference between simulated and measured realized gain for the PEWAN-A sample across all measurement configurations.}
\label{fig:3methods_2} 
\end{figure}
\begin{figure}
\centering
\includegraphics[]{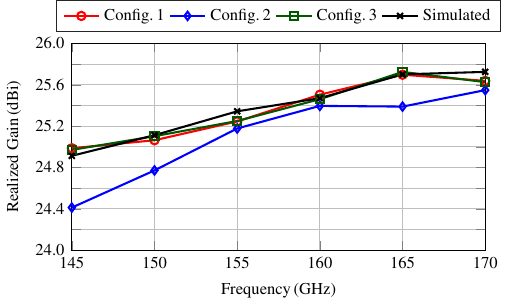}
\caption{Comparing simulated and measured realized gain at each operational frequency point for the PEWAN-A sample in cluster\,2 across all three measurement configurations.}
\label{fig:3methods_3} 
\end{figure}

In order to avoid repetitive plots, Fig.~\ref{fig:3methods_2} illustrates a comparison of the absolute average difference between measured and simulated realized gain value ($|\Delta G_{\text{avg}}|$) in the overlapping frequency range obtained by three different measurement configurations for the PEWAN-A sample across all clusters. As expected for the second measurement configuration, there is a slightly larger deviation in the absolute average value between the simulated and realized gain. This is in part due to the usage of the two FLANN samples with the WR6.5 frequency extender which would introduce some loss due to mismatch and the reasoning mentioned in Section~\ref{sec2} due to wider aperture area of the PEWAN1028 model. It is also observed that at cluster\,2, $|\Delta G_{\text{avg}}|$ is very close for the first and third measurement configurations, thereby highlighting the importance of this measurement distance when choosing appropriate AUT combinations.
In Fig.~\ref{fig:3methods_3} the measured realized gain obtained at each operational frequency point  for the PEWAN-A sample in cluster\,2 across all three measurement configurations are compared against the respective simulated values. It is observed that the values obtained from the first and the third measurement configurations are comparable not only to the simulated value with minimal deviations but also with one another while following a similar trend. This points out to the efficacy of utilizing frequency extenders as a third AUT option thereby relieving the factor of inventory constraints in accurate gain measurements. As for the second measurement configuration, noteable discrepancies occur at certain frequency points justifying the reasoning of mismatch between the FLANN samples and the WR6.5 frequency extender can have poorer match near the edges of the band which may produce amplitude errors that effect the three-antenna algebra.
\subsection{Comparison with Extrapolation Technique}

In order to determine the validity of the CCM technique proposed in this article, the  obtained results are compared against the classical extrapolation technique as described in \cite{3}. Extrapolation-based gain is usually determined from distance sweep measurements performed over a wide separation range which usually begins in the NF of the antenna and is extrapolated to give the infinite range FF gain.  

To comply with the fundamental criteria presented in \cite{3}, we performed extrapolation measurements across a separation range of 35\,cm to 175\,cm between the antenna apertures at a 0.2\,mm step size to achieve $d_\text{max}/d_\text{min}$ $\geq$ 3. This provided us a ratio of 5:1 which improves the conditioning of the least-squares fit. As the same setup is employed to perform extrapolation measurements, the limitation of the linear rail system was tackled by measuring at several clusters of distances within the separation range. To ensure measurement data integrity, overlapping measurements between clusters were acquired as well to build a stitched dataset in order to perform the extrapolation fit. 
\begin{figure}[!t]
\centering
\begin{tabular}[b]{c}
\includegraphics[]{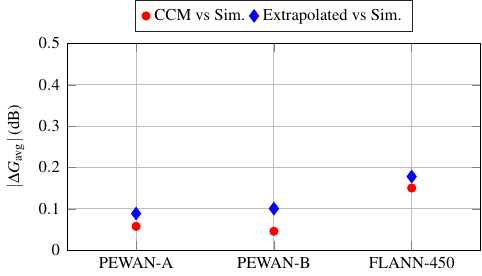} \\
\footnotesize (a) First Measurement Configuration
\end{tabular} \qquad
\begin{tabular}[b]{c}
\includegraphics[]{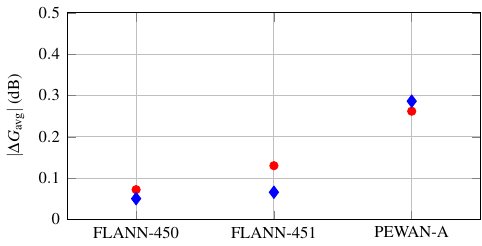} \\
\footnotesize (b) Second Measurement Configuration
\end{tabular}
\caption{Absolute average difference between simulated and measured realized gain of all AUTs for the first and second measurement configurations.}
\label{fig:extrapolation}
\end{figure}
It is important to note that the requirements for extrapolation measurements given in \cite{3} are more of a recommendation than strict guidelines, especially because it was made clear that the requirements were developed at that time based on the antennas being studied. The upper bound specified therein is motivated by the practical risk of ground reflections for larger ranges. However, since our table-top setup is compact and packed with high performance absorbers we are actively reducing the effects of this concern and are able to safely extend $d_\text{max}$ without corrupting the data. Another important reason is that the antennas used in \cite{3} are all of similar physical sizes. Another example of such an analogous measurement process has been presented at W-band in \cite{zsong}. Because our AUTs contained a mix of physically larger 110--170\,GHz horns and relatively smaller 145--220\,GHz horn, we chose a slightly longer separation span in order to give the extrapolation fit more leverage in terms of $1/d$ for all three antenna pair combinations.

Measurement results obtained using the extrapolation technique have been illustrated in Fig.~\ref{fig:extrapolation}. It presents the absolute average offset between measured and simulated realized gain values ($|\Delta G_{\text{avg}}|$) obtained from cluster\,2 using the CCM technique and the extrapolation measurement technique. In Fig.~\ref{fig:extrapolation}(a), the obtained values for the PEWAN-A sample are 0.06\,dB and 0.08\,dB using the CCM and extrapolation techniques respectively and similarly for the PEWAN-B sample the values obtained are 0.05\,dB and 0.10\,dB respectively. Minor differences in values between the specimens can be accounted to tolerances from model fabrications as well as measurement uncertainties arising from performing a stitched extrapolation as stated previously.

In the case of the FLANN-450 sample, the values obtained are 0.15\,dB and 0.18\,dB for the CCM and extrapolation techniques respectively. These numbers obtained are well within the standards considering the measurement setup. In Fig.~\ref{fig:extrapolation}(b), the results obtained for the second measurement configuration are presented. The $|\Delta G_{\text{avg}}|$ values obtained for the CCM technique are 0.07\,dB and 0.13\,dB for the FLANN-450 and FLANN-451 samples respectively. The values obtained by the extrapolation technique are closely matching with 0.05\,dB and 0.06\,dB for the FLANN-450 and FLANN-451 samples respectively. The $|\Delta G_{\text{avg}}|$ values obtained for the PEWAN-A sample are 0.26\,dB and 0.28\,dB between the CCM and extrapolation techniques respectively. 
 
Here, a case can be made for the compact cluster measurement technique based on the measurement results mentioned above. There are two important considerations to take into account while using the extrapolation approach in \cite{3}. Firstly, averaging out the standing wave ripples in distance originating mainly from ground reflections and the measurement range and secondly, de-embedding of the finite distance coupling via the $1/d$ expansion. The CCM technique accounts for the first factor by having a dense sampling over a local window and multiple averaged measurement runs to suppress the ripples. Furthermore, the absorber layouts and positioning in order to compensate for any reflections have also been taken care of in the measurement setup. In case of the second factor, since this measurement setup is inherently in the FF considering the effects of both antennas using \eqref{eq:revised_FF}, a simple Friis model is close to the asymptotic behaviour making the $1/d$ correction from the extrapolation technique small enough to be safely ignored. At such separation distances, the Friis equation alone and local averaging can be mathematically equivalent.

\begin{figure}
\centering
\includegraphics[]{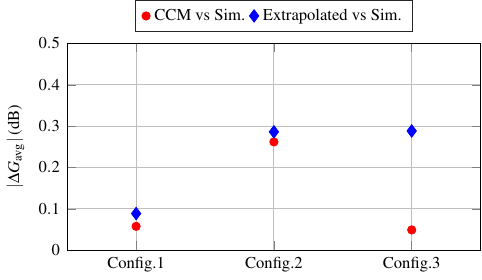}
\caption{Absolute average difference between simulated and measured realized gain for the PEWAN-A sample across all measurement configurations.}
\label{fig:extrapol3} 
\end{figure}

In Fig.~\ref{fig:extrapol3}, we notice the difference between the absolute averaged measured and simulated gain values for the PEWAN-A sample across all three measurement configurations for the two measurement techniques. The values obtained for the CCM technique are 0.06\,dB, 0.26\,dB and 0.05\,dB for the first, second and third measurement configurations respectively. The values obtained for the extrapolation technique are 0.08\,dB, 0.28\,dB and 0.29\,dB for the first, second and third measurement configurations respectively. The slightly higher values of the second configuration in both techniques are caused by, as previously mentioned, the use of mismatched antenna combinations. Nevertheless, the results thus confirm the accuracy of the compact cluster method with respect to the standard extrapolation and also the harmony between both techniques.

For these reasons, the proposed CCM method can be considered as an alternative to the standard extrapolation technique as the physics of the two methods are consistent, and the new approach provides an opportunity to measure antennas in compact spaces, at a faster measurement acquisition time, and for facilities which may not have long measurement ranges or with inventory constraints.

\section{Conclusion} \label{conclusion}
This article has presented measured gain results based on the proposed modified far-field formulation which considers the combined effects of both AUTs. The effectiveness of this approach has been validated for gain characterization in the sub-THz range of 145--170\,GHz for various antenna pair configurations. The proposed modified theoretical model matched practical measurements with reasonable accuracy, enabling a compact cluster based measurement setup while considering the effects of both AUTs without the need of a dedicated anechoic chamber. Despite the use of mismatched instrumentation combinations, the extracted gain values remained within acceptable limits for the industry standards, demonstrating the flexibility of the measurement setup. This suggests that even in the cases where an ideal AUT is unavailable due to cost constraints, the open-ended waveguide flange of the frequency extenders may be utilized as the third AUT to extract meaningful and reliable gain measurement results. The findings highlight the practicality of this method offering a cost-effective and compact solution for antenna characterization at sub-THz frequency ranges. Future work includes focus on further refining the measurement setup to account for additional systematic uncertainties, repeatabilities, and the need of any proximity corrections. It also involves the exploration of its applicability to a wider range of antenna designs, such as, lens antenna designs and measurement frequency ranges.

\appendices
\counterwithin*{equation}{section}
\renewcommand\theequation{\thesection.\arabic{equation}}
\section{Mathematical Approximation for the Revised Far-Field Formulation}\label{ff_approximation}
The aim is to approximate the expression in \eqref{eq:fourth_oder_distance} to \eqref{eq:revised_FF}. First, let the expression in \eqref{eq:fourth_oder_distance} be rewritten as follows
\begin{equation}
d_\text{ff} = \frac{2}{\lambda} \sqrt{(D_1^2)^2 + (D_2^2)^2}. 
\label{eq:a1}
\end{equation}
The mathematical approximation for square root of sums is given as
\begin{equation}
\sqrt{a^2 + b^2} \approx a+b 
\label{eq:a2}
\end{equation}
when $a$ $\approx$ $b$ or one of them dominates. Approximation for both cases have been thus presented.\\
$Case\,1$: When $a$ $\gg$ $b$\\
Factorizing the expression in \eqref{eq:a2} gives
\begin{equation}
\sqrt{a^2 + b^2} = \sqrt{a^2 \left(1 + \frac{b^2}{a^2} \right)} = a \sqrt{1 + x}
\label{eq:a3}
\end{equation}\\
where $x = (b/a)^2$.
The binomial first order approximation from Taylor series expansion of a square root function is given as
\begin{equation}
\sqrt{1 + x} \approx 1+\frac{x}{2}. 
\label{eq:a4}
\end{equation}\\
Applying the binomial approximation from \eqref{eq:a4} to \eqref{eq:a3} gives:
\begin{equation}
\sqrt{a^2 + b^2} \approx {a\left(1 + \frac{b^2}{2a^2} \right)} \approx \left(a + \frac{b^2}{2a} \right).
\label{eq:a5}
\end{equation}\\
For practical applications, we can say that ${b^2}/{2a} \ll b$ and hence $\sqrt{a^2 + b^2} \approx a+b$ would be a safe approximation. Relating this to \eqref{eq:a1} would then give us the formulation of the revised FF distance as expressed in \eqref{eq:revised_FF}.\\
$Case\,2$: When $a$ $\approx$ $b$\\
A small perturbation parameter $z = {b}/{a}$, where $z$ $\approx$ 1 is defined. Hence \eqref{eq:a3} can be written in this case as:
\begin{equation}
\sqrt{a^2 + b^2} = \sqrt{a^2 \left(1 + \frac{b^2}{a^2} \right)} = a \sqrt{1 + z^2}.
\label{eq:a6}
\end{equation}\\
Considering that small deviations can be expected from unity, $\sqrt{1 + z^2}$ is expanded using Taylor series around $z$ = 1 as
\begin{equation}
\sqrt{1 + z^2} = 1 + \frac{z^2 - 1}{2} .
\label{eq:a7}
\end{equation}
The expansion is limited to the first order approximation as that would capture the dominant behaviour for small perturbations. Higher order correction terms would keep getting smaller, thereby contribute significantly little to the accuracy, and add unnecessary complexity to derive the analytical expression.\\
Substituting \eqref{eq:a7} in \eqref{eq:a6} gives
\begin{equation}
\sqrt{a^2 + b^2} \approx a + \frac{b^2 - a^2}{2a}.
\label{eq:a8}
\end{equation}\\
Applying an algebraic identity on the ${(b^2 - a^2)}$ term gives
\begin{equation}
\sqrt{a^2 + b^2} \approx a + \frac{(b - a)(b+a)}{2a}.
\label{eq:a9}
\end{equation}\\
Using an arithmetic mean approximation on the $(b+a)$ term implies that $b+a$ $\approx$ 2$a$. Thus 
\begin{equation}
\sqrt{a^2 + b^2} \approx a + (b - a).
\label{eq:a10}
\end{equation}\\
Since we assume $a$ $\approx$ $b$, another interchangeable equivalent solution could be reached as
\begin{equation}
\sqrt{a^2 + b^2} \approx b + (a - b).
\label{eq:a11}
\end{equation}\\
Both \eqref{eq:a10} and \eqref{eq:a11} would approximate the square root differently leaning to favour one variable over the other. Since neither is a perfect approximation, converging a first order solution of the perturbation series using summation methods affords the possibility to provide a good approximation. This technique of improving accuracy is a commonly used principle in perturbation methods, asymptotic expansions, and numerical techniques in first order approximations~\cite[\S7.1]{math_book}.
Following this, averaging the two approximations would lead to give $\sqrt{a^2 + b^2} \approx a+b$ and hence the FF formulation in \eqref{eq:revised_FF} would be justified in this case as well.

\section{Calculation of the Standard Deviation From Measurement Results}\label{std_dev}
The averaged gain value at each measurement point \( m \) and frequency point \( f \) is given by
\begin{equation}
\bar{G}_{m,f} = \frac{1}{N} \sum_{r=1}^N G_{m,f}^{(r)}.
\label{eq:b1}
\end{equation}
where  \( r \) represents the repetition index, implying the number of the  repeated measurements that have been carried out. This forms Dataset A, containing the averaged gain values in dB for all \( m \) and \( f \) values. 

Using Dataset A, the gain at each measurement point \( m \) and frequency point \( f \) is calculated using the three-antenna technique, and the resulting Friis transmission equation is denoted as \( G_{m,f}^{\text{calc}} \). This forms Dataset B, which contains the calculated gain values in dB. 

To assess the variation in the calculated gain values in Dataset B, standard deviation is applied across all the measurement points \( m \) at each frequency point \( f \). The standard deviation \( \sigma_f \) for the calculated gain at frequency point \( f \) is given by
\begin{equation}
\sigma_f = \sqrt{\frac{1}{M} \sum_{m=1}^M \left(G_{m,f}^{\text{calc}} - \bar{G}_f^{\text{calc}}\right)^2}
\end{equation}
where $M$ is the total number of measurement points.
\( \bar{G}_f^{\text{calc}} \) is the average calculated gain over all measurement points for frequency $f$ as
\begin{equation}
\bar{G}_f^{\text{calc}} = \frac{1}{M} \sum_{m=1}^M G_{m,f}^{\text{calc}}.   
\end{equation}  
This standard deviation provides a measure of how much the calculated gain varies across measurement points at each frequency point.

\section*{Acknowledgments}
The authors would like to thank Mr. Markku Jokinen for help with measurement data acquisition. Keysight Inc. has supported the research by donating measurement equipment. Simulation results have been obtained using computational resources from CSC--IT Center for Science, Finland.

\bibliographystyle{IEEEtran}
\bibliography{IEEEabrv,bibliography}

@STRING{IEEE_J_PROC       = "Proc. {IEEE}"}

@STRING{IEEE_J_AP         = "{IEEE} Trans. Antennas Propag."}

@STRING{IEEE_J_TTHZ       = "{IEEE} Trans. {THz} Sci. Technol."}

@STRING{IEEE_J_IM         = "{IEEE} Trans. Instrum. Meas."}

@IEEEtranBSTCTL{modify_bib,
ctluse_forced_etal = {yes},
ctlmax_names_forced_etal={6},
ctlnames_show_etal = {1},
}

@MISC{1, author={P{\"a}rssinen, Aarno and Alouini, Mohammad and Berg, Markus and Kuerner, Thomas and Ky{\"o}sti, Pekka and Leinonen, Marko E. and Matinmikko-Blue, Marja and McCune, Earl and Pfeiffer, Ullrich and {Wambacq, Piet (Eds.)}}, title={White Paper on {RF} Enabling {6G} -- Opportunities and Challenges from Technology to Spectrum}, howpublished={University of Oulu}, year={2020}, url={http://urn.fi/urn:isbn:9789526228419}, note={{White} paper, (6G Research Visions, No. 13)}}

@book{2,
author = {Eichler, Taro and Ziegler, Robert},
year = {2022},
month = {July},
pages = {},
publisher = {Rohde \& Schwarz}, 
title = {White paper: Fundamentals of THz technology for 6G},
url = {https://www.rohde-schwarz.com/solutions/test-and-measurement/wireless-communication/cellular-standards/6g/white-paper-fundamentals-of-thz-technology-for-6g-by-rohde-schwarz-registration_255934.html}
}

@ARTICLE{3,
  author={Newell, A. and Baird, R. and Wacker, P.},
  journal=IEEE_J_AP, 
  title={Accurate measurement of antenna gain and polarization at reduced distances by an extrapolation technique}, 
  year={1973},
  volume={21},
  number={4},
  pages={418-431},
  month={Apr.},
  doi={10.1109/TAP.1973.1140519}}

@ARTICLE{4,
  author={Repjar, Andrew G. and Newell, Allen C. and Baird, Ramon C.},
  journal=IEEE_J_IM, 
  title={Antenna Gain Measurements by an Extended Version of the {NBS} Extrapolation Method}, 
  year={1983},
  volume={32},
  number={1},
  pages={88-91},
  month={Mar.},
  doi={10.1109/TIM.1983.4315016}}

@ARTICLE{5,
  author={Friis, H.T.},
  journal={Proc. IRE}, 
  title={A Note on a Simple Transmission Formula}, 
  year={1946},
  volume={34},
  number={5},
  month={May},
  pages={254-256},
  doi={10.1109/JRPROC.1946.234568}}

@ARTICLE{6,
  author={},
  journal={ANSI/IEEE Std 149-1979}, 
  title={{IEEE} Standard Test Procedures for Antennas}, 
  year={1979},
  volume={},
  number={},
  pages={78-98},
  doi={10.1109/IEEESTD.1979.120310}}

@INPROCEEDINGS{7,
  author={Xiao, Liu and Donglin, Meng and Pan, Huang and Zhenfei, Song and Jinyuan, Li and Haoyu, Lin},
  booktitle={2018 11th Global Symp. Millim. Waves (GSMM)}, 
  title={Accurate Measurement for Millimeter-wave Antenna Based on the Extrapolation Range}, 
  year={2018},
  volume={},
  number={},
  pages={1-3},
  month={May},
  address={Boulder, CO, USA},
  doi={10.1109/GSMM.2018.8439457}}

@inproceedings{8,
author = {Zhao, Xing and Ban, Hao and Wang, Rongcai and Liu, Xiao},
  booktitle={2023 IEEE 7th Int. Symp. Electromagn. Compat. (ISEMC)}, 
year = {2023},
month = {Oct.},
pages = {1-4},
address={Hangzhou, China},
title = {Standard antenna measurement for mobile communication based on the three-antenna extrapolation technique},
doi = {10.1109/ISEMC58300.2023.10370312}
}

@techreport{8i,
  author      = "",
  title       = "Measurement method of passive antenna for mobile communication system",
  institution = "Ministry of Industry and Information Technology of the People's Republic of China",
  year        = "2020",
  type        = "Industry Standard:",
  number      = "YD/T 2868-2020",
  address     = "Beijing, China",
  month       = "",
  note        = ""
}

@ARTICLE{9,
  author={Bowman, R.R.},
  journal=IEEE_J_PROC, 
  title={Field strength above 1 {GHz}: Measurement procedures for standard antennas}, 
  year={1967},
  volume={55},
  number={6},
  pages={981-990},
  month={June},
  doi={10.1109/PROC.1967.5712}}

@INPROCEEDINGS{12,
  author={Katsushige Harima and Makoto Sakasai and Katsumi Fujii},
  booktitle={2008 IEEE Int. Symp. Electromagn. Compat. (ISEMC)}, 
  title={Determination of gain for pyramidal-horn antenna on basis of phase center location}, 
  year={2008},
  volume={},
  number={},
  pages={1-5},
  month={Aug.},
  address={Detroit, MI, USA},
  doi={10.1109/ISEMC.2008.4652010}}

@INPROCEEDINGS{13,
  author={Hertel, T.W.},
  booktitle={IEEE Ant. Propag. Soc. Int. Symp. Digest. (APS/URSI)}, 
  title={Phase center measurements based on the three-antenna method}, 
  year={2003},
  volume={3},
  number={},
  pages={816-819},
  month={June},
  address={Columbus, OH, USA},
  doi={10.1109/APS.2003.1220035}}

@ARTICLE{14,
  author={Hollmann, H.},
  journal=IEEE_J_IM, 
  title={Accurate gain measurement of horn antennas in the shortened far field}, 
  year={1989},
  volume={38},
  number={2},
  pages={617-618},
  month={Apr.},
  doi={10.1109/19.192361}}

@ARTICLE{15,
  author={van den Biggelaar, A. J. and Geluk, S. J. and Jamroz, B. F. and Williams, D. F. and Smolders, A. B. and Johannsen, U. and Bronckers, L. A.},
  journal=IEEE_J_AP, 
  title={Accurate Gain Measurement Technique for Limited Antenna Separations}, 
  year={2021},
  volume={69},
  number={10},
  pages={6772-6782},
  month={Oct.},
  doi={10.1109/TAP.2021.3069583}}

@ARTICLE{16,
  author={Rivera-Lavado, Alejandro and Ali, Muhsin and Gallego-Cabo, Daniel and García-Mu\~{n}oz, Luis-Enrique and Lioubtchenko, Dmitri V. and Carpintero, Guillermo},
  journal=IEEE_J_TTHZ, 
  title={Contactless {RF} Probe Interconnect Technology Enabling Broadband Testing to the Terahertz Range}, 
  year={2023},
  volume={13},
  number={1},
  month={Jan.},
  pages={34-43},
  doi={10.1109/TTHZ.2022.3213470}}

@ARTICLE{uno,
  author={Uno, T. and Adachi, S.},
  journal=IEEE_J_AP, 
  title={Range distance requirements for large antenna measurements}, 
  year={1989},
  volume={37},
  number={6},
  month={June},
  pages={707-720},
  doi={10.1109/8.29357}}

@techreport{18,
  author      = "William T. Slayton",
  title       = "Design and Calibration of Microwave Antenna Gain Standards",
  institution = "Naval Research Laboratory",
  year        = "1954",
  type        = "Technical Report",
  number      = "4433",
  address     = "Washington D.C., USA",
  month       = "Nov.",
  note        = ""
}

@TechReport{3gpp, 
author = "3GPP",
title = "{5G}; {NR}; {D}erivation of test tolerances and measurement uncertianity for {U}ser {E}quipment {(UE)} and conformance test cases (Release 15)", 
institution = "3GPP", 
year = "2019",
number = "TR 138.903 V15.2.0 (2019-04)" }

@ARTICLE{intro1,
  author={Rahmat-Samii, Y. and Galindo-Israel, V. and Mittra, R.},
  journal=IEEE_J_AP, 
  title={A plane-polar approach for far-field construction from near-field measurements}, 
  year={1980},
  volume={28},
  number={2},
  pages={216-230}}

@ARTICLE{intro2,
  author={Hamid, M.},
  journal=IEEE_J_AP, 
  title={The radiation pattern of an antenna from near-field correlation measurements}, 
  year={1968},
  volume={16},
  number={3},
  pages={351-353}}

@ARTICLE{intro3,
  author={Lonnqvist, A. and Koskinen, T. and Hakli, J. and Saily, J. and Ala-Laurinaho, J. and Mallat, J. and Viikari, V. and Tuovinen, J. and Raisanen, A.V.},
  journal=IEEE_J_AP, 
  title={Hologram-based compact range for submillimeter-wave antenna testing}, 
  year={2005},
  volume={53},
  number={10},
  pages={3151-3159},
  doi={10.1109/TAP.2005.856344}}

@ARTICLE{intro4,
  author={Hakli, J. and Koskinen, T. and Lonnqvist, A. and Saily, J. and Viikari, V. and Mallat, J. and Ala-Laurinaho, J. and Tuovinen, J. and Raisanen, A.V.},
  journal=IEEE_J_AP, 
  title={Testing of a 1.5-m reflector antenna at 322 GHz in a CATR based on a hologram}, 
  year={2005},
  volume={53},
  number={10},
  pages={3142-3150},
  doi={10.1109/TAP.2005.856343}}

@ARTICLE{gordon,
  author={Mayhew-Ridgers, G. and Odendaal, J.W. and Joubert, J.},
  journal=IEEE_J_IM, 
  title={Horn antenna analysis as applied to the evaluation of the gain-transfer method}, 
  year={2000},
  volume={49},
  number={5},
  month={Oct.},
  pages={949-958},
  doi={10.1109/19.872913}}

@inproceedings{Pivnenko,
  title={Fast and accurate measurement of on-axis gain and on-axis polarization at a finite distance},
  author={Sergey Pivnenko and Olav Breinbjerg},
  booktitle={2013 7th Eur. Conf. Antennas Propag. (EuCAP)},
  year={2013},
  pages={810-814},
  address={Gothenburg, Sweden},
  month={Apr.}
}

@ARTICLE{odendaal,
  author={Odendaal, J.W. and Pistorius, C.W.I.},
  journal=IEEE_J_IM, 
  title={A method to measure the aperture field and experimentally determine the near-field phase center of a horn antenna}, 
  year={1993},
  volume={42},
  number={1}, 
  month={Feb.},
  pages={51-53},
  doi={10.1109/19.206680}}

@book{balanis,
title = {Antenna Theory: Analysis and Design, 3rd ed.},
author = {C. A. Balanis},
publisher = {John Wiley},
address={Hoboken, NJ},
year = {2005},
url = {}}

@book{math_book,
author = {Bender, Carl and Orszag, Steven},
year = {1978},
publisher = {McGraw-Hill},
title = {Advanced Mathematical Methods for Scientists and Engineers: Asymptotic Methods and Perturbation Theory},
volume = {},
}

@ARTICLE{morgan,
  author={Hunter, John D. and Morgan, I. G.},
  journal=IEEE_J_IM, 
  title={Near-Field Gain Correction for Transmission between Horn Antennas}, 
  year={1977},
  volume={26},
  number={1},
  month={Mar.},
  pages={58-61},
  doi={10.1109/TIM.1977.4314485}}

@ARTICLE{wband_comp,
  author={Kang, Jin-Seob and Kang, No-Weon and Gentle, David G. and MacReynolds, Katherine and Francis, Michael H.},
  journal=IEEE_J_IM, 
  title={Intercomparison of Standard Gain Horn Antennas at {$W$} -Band}, 
  year={2011},
  volume={60},
  number={7},
  month={July},
  pages={2627-2633},
  doi={10.1109/TIM.2010.2103413}}

@techreport{military,
  author      = "",
  title       = {{R}adio frequency spectrum characteristics, measurement of},
  institution = "",
  year        = "1973",
  type        = "Military Standard:",
  number      = "MIL-STD-449D",
  address     = "",
  month       = "February",
  note        = ""
}

@INPROCEEDINGS{zsong,
  author={Song, Zhenfei and Gentle, David and Lin, Haoyu and Liu, Xiao and Li, Jinyuan},
  booktitle={IET Int. Radar Conf. 2015}, 
  title={Accurate gain calibration for a {WR10} standard gain horn {(SGH)} using the three-antenna extrapolation technique}, 
  year={2015},
  volume={},
  month={Oct.},
  address={Hangzhou, China},
  number={},
  pages={1-5},
  doi={10.1049/cp.2015.1270}}

@Manual{absorber,
  title        = {WAVASORB\textsuperscript{\tiny\textregistered} VHP Advanced Broadband Pyramidal Absorber},
  year         = {2024},
  organization = {Emerson \& Cuming Anechoic Chambers nv},
  address={Westerlo, Belgium},
  url          = {https://www.ecanechoicchambers.com/pdf/WAVASORB%20VHP%20datasheet%20full%20version_final.pdf},
}

@ARTICLE{Hirano2014,
  author={Hirano, Takuichi and Hirokawa, Jiro and Ando, Makoto},
  journal=IEEE_J_AP, 
  title={Errors in Shortened Far-Field Gain Measurement Due to Mutual Coupling}, 
  year={2014},
  volume={62},
  number={10},
  month={Oct.},
  pages={5386-5388},
  doi={10.1109/TAP.2014.2342757}}

@ARTICLE{jaarsveld,
  author={Mayhew-Ridgers, Gordon and van Jaarsveld, Paul A. and Odendaal, Johann W.},
  journal=IEEE_J_AP, 
  title={Three-Antenna Characterization Techniques Employing Spherical Near-Field Scanning With Higher-Order Probe Correction}, 
  year={2023},
  volume={71},
  number={9},
  pages={7220-7228},
  doi={10.1109/TAP.2023.3295492}}
\end{document}